\documentclass[aps,prl,twocolumn,showpacs]{revtex4}

\usepackage{epsfig}
\usepackage{amsbsy,latexsym}
\usepackage{amsmath}
\usepackage{amssymb, mathrsfs}
\usepackage[mathscr]{eucal}
\usepackage{hyperref}

\def\d{\operatorname{d}}
\def\<{\langle}
\def\>{\rangle}
\def\Tr{\operatorname{Tr}}

\def\:{\hbox{\bf :}}

\def\set#1{{\sf #1}}
\def\dag{\dagger}
\def\geq{\geqslant}
\def\leq{\leqslant}

\def\sH{\mathcal{H}}
\def\sK{\mathcal{K}}

\def\Lin{\mathcal{L}}

\def\supp{\set{Supp}}
\def\qed{$\,\blacksquare$\par}
\def\kk{\rangle\!\rangle}
\def\bb{\langle\!\langle}

\newcommand{\ket}[1]{| #1 \rangle}
\newcommand{\bra}[1]{\langle #1 |}
\newcommand{\Ket}[1]{| #1 \rangle \! \rangle}
\newcommand{\Bra}[1]{\langle \! \langle #1 |}

\newcommand{\KetBra}[2]{\Ket{#1} \Bra{#2}}

\newcommand{\ketbra}[2]{\ket{#1} \bra{#2}}
\newcommand{\hilb}[1]{\mathcal{#1}}

\newtheorem{lemma}{Lemma}

\newtheorem{theorem}{Theorem}
\def\Proof{\medskip\par\noindent{\bf Proof. }} 

\begin{document}
\title{Minimal computational-space implementation of multi-round quantum protocols}

 \author{Alessandro Bisio}
 \affiliation{QUIT group,  Dipartimento di Fisica ``A. Volta'', INFN Sezione di Pavia, via Bassi
   6, 27100 Pavia, Italy.} 
 \author{Giulio Chiribella} 
\affiliation{Perimeter Institute for Theoretical Physics, 31 Caroline St. North, Waterloo, Ontario N2L 2Y5, Canada } 
 \author{Giacomo Mauro D'Ariano}
 \affiliation{QUIT group,  Dipartimento di Fisica ``A. Volta'', INFN Sezione di Pavia, via Bassi
   6, 27100 Pavia, Italy.} 
 \author{Paolo Perinotti} 
 \affiliation{QUIT group,  Dipartimento di Fisica ``A. Volta'', INFN Sezione di Pavia, via Bassi
   6, 27100 Pavia, Italy.} 
 \date{ \today}
 \begin{abstract} 

   A single-party strategy in a multi-round quantum protocol can be
   implemented by sequential networks of quantum operations connected
   by internal memories. Here provide the most efficient realization
   in terms of computational-space resources.

\end{abstract}
\maketitle


Many results in Quantum Information \cite{Nielsen} and
 Quantum Estimation Theory \cite{holevo,helstrom}  have
been achieved through the general description of states and
measurements in terms of density matrices and positive operator-valued measures (POVM's), respectively.
 The advantages of this formalism are evident in optimization
tasks, like  e.g. state discrimination, where one can look for the optimal measurement without considering the specific details of the apparatus. Furthermore, the optimization of preparation/measurement devices  is reduced to the optimization of positive operators, for which many powerful techniques are known.     Similar advantages are provided by the description of physical transformations as quantum channels (completely positive trace-preserving maps), which  in turn  can be represented by positive operators via the
Choi-Jamio\l kowski isomorphism \cite{choijam}.  

The usage of the Choi-Jamio\l kowski isomorphism is well established
in quantum information theory \cite{schum,cir} since the early works
on ancilla-assisted tomography \cite{darlop,leung}.  Recently, the
Choi-Jamio\l kowski representation has been extended to more complex
quantum devices, consisting of sequences of channels, quantum
operations and POVM's connected by internal wires
\cite{watrousgame,architecture, comblong}. In particular, Ref.
\cite{watrousgame} considered the application of these sequential
networks to represent single-party strategies in multi-round quantum
games, while Refs. \cite{architecture,comblong} showed how these
networks can implement a variety of higher-order quantum information
processing tasks, such as transforming states into channels, channels
into channels, and even networks into networks.  Refs.
\cite{architecture,comblong} also coined the name \emph{quantum combs}
for the Choi-Jamio\l kowski operators associated to sequential
networks, and developed a simple set of rules to describe the
interlinking of networks in terms of the corresponding operators.  In
this framework, once a specific task is fixed (e.g.  cloning a channel
\cite{cloning} or estimating the POVM of a detector \cite{plenio}) one
can search for the quantum protocol that optimally realizes it.
Having a simple description now becomes indispensable: since a quantum
protocol is implemented by a complex network of devices, optimizing
each device separately is not a viable approach.
In the new framework, instead, 
the optimization of the protocol  is reduced to the  optimization of a single positive operator subject to linear constraints. In the simplest cases the search can be also implemented automatically through matlab routines \cite{boyd,watgut}.

Once the optimal Choi-Jamio\l kowski operator has been found, however,
one needs a way to unzip the information contained in it and to find a
physical implementation of the network.
In this Letter we solve this problem, presenting an automatic
procedure that, given the Choi-Jamio\l kowski operator of a quantum
network, allows to construct a concrete implementation of it as a
sequence of elementary devices.  Among all possible implementations,
the present one minimizes the computational space, that is, at each
step it uses the smallest possible dimension of the Hilbert spaces.
Our procedure can be fully automatized in a computer software,
accepting as an input the Choi-Jamio\l kowski representation of the
network and providing as an output the matrix representation of the
operations that must be performed at each stage of the protocol.
After the operations in the network have been determined one can look
for a further decomposition of them into elementary gates, using e.g.
the techniques of Refs.  \cite{salomaa, shende}.


We now review the basic concepts and results of the general
theory of quantum networks as presented in Refs.
\cite{architecture,comblong}.
\begin{figure}[t]
 \includegraphics[width=\columnwidth]{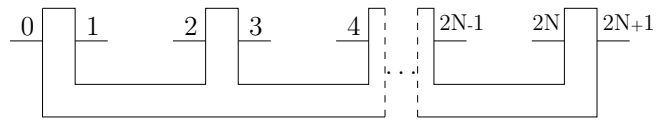}
  \caption{\label{fig:comb} A quantum comb with $N$ slots.
    Information flows from left to right.  The causal structure of the
    comb implies that the input system $m$ cannot influence the output
    system $n$ if $m>n$.}
\end{figure}
\begin{figure}[t]
  \includegraphics[width=\columnwidth]{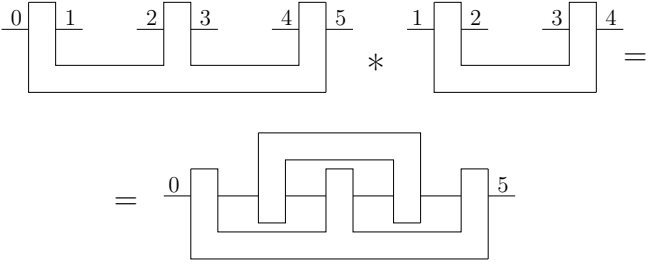}
\caption{\label{fig:comblink} Linking of two combs.  We identify the
  wires with the same label.}
\end{figure}
The most general quantum device is
a quantum circuit board, namely a network of
quantum devices with open slots  to which variable sub-circuits can be
linked.
By stretching and rearranging the internal wires of the network,
 we can give to each quantum circuit board the shape of a comb, like in Fig. \ref{fig:comb}.
 The empty slots of the circuit board
become the empty spaces between two teeth of the comb.
 Referring to Fig. \ref{fig:comb}, each wire is labeled with a natural number,
 which is even for the input wires and odd for the output ones;
 the
corresponding Hilbert spaces are
labelled in the same fashion (that is, the Hilbert space of the system
 represented by the wire $i$ is denoted by  $\hilb{H}_i)$).
 The ordering of the slots results from the causal ordering defined by the flow of quantum information
from input to output; with our notation we have 
that input system in wire $i$ can influence the output system in a wire $j>i$
 but not in a wire $k<i$. 
Two circuit boards $\mathscr C_1$
and $\mathscr C_2$ can be connected by linking some outputs of $\mathscr C_1$ with
inputs of $\mathscr C_2$, thus forming a new board $\mathscr C_3 :=\mathscr  C_1 * \mathscr C_2$. We adopt
the convention that wires that are connected are identified by the
same label (see Fig.  \ref{fig:comblink}).

In the following we will often use the isomorphism between linear operators in ${\rm Lin}(\hilb{H})$
and states in $\hilb{H} \otimes \hilb{H}$:
\begin{align} \label{doubleketeq}
&A = \sum_{nm}\bra{n}A\ket{m} \ketbra{n}{m} 
\leftrightarrow
\Ket{A} = \sum_{nm}\bra{n}A\ket{m} \ket{n}\ket{m}& \nonumber
\end{align}
where $\{ \ket{n} \}$ is a fixed orthonormal basis. 

The {\em quantum comb} $C$ associated to a circuit board  $\mathscr C$ with $N$
input/output systems is the Choi-Jamio\l kowski operator of the multipartite
channel representing the input/output transformation that
the board performs from states on  $\sH_{\rm{in}} := \bigotimes_{j =
  0}^{N-1} \sH_{2j}$ to states on $\sH_{\rm{out}} :=
\bigotimes_{j=0}^{N-1} \sH_{2j+1}$,  $\sH_n$ being the Hilbert space of the $n$-th
system.
A quantum comb is then a positive operator acting on $\sH_{\rm{out}} \otimes \sH_{\rm{in}}$
and it is defined as follows:
\begin{equation}
C_{\rm{out} \,  \rm{in}} := (\mathscr C \otimes {\mathcal{I}_{\rm{in}}})(\KetBra{I}{I}_{\rm{in}\, \rm{in}})
\end{equation}
(for clarity here we use the notation 
$\hilb{H}_{ab} \equiv  \hilb{H}_a \otimes \hilb{H}_b $, $A_{ab}$ to mean $A  \in {\rm Lin}(\hilb{H}_{ab})$, 
$\ket{\psi}_{b}$ to mean $\ket{\psi} \in \hilb{H}_b$, and $\Ket{A}_{ab}$ to mean 
$\Ket{A} \in \hilb{H}_{ab}$).
 It can be proved that the causal structure is equivalent to the recursive
normalization condition
\begin{equation}\label{recnorm}
\Tr_{2k-1} [ C^{(k)}] = I_{2k-2} \otimes C^{(k-1)} \qquad k=1, \dots, N~
\end{equation} 
where $C^{(N)}=C$, $C^{(0)} =1$, $C^{(k)}
\in\Lin(\sH_{\rm{out}_k} \otimes \sH_{\rm{in}_k})$
with
 $\sH_{\rm{in}_k} = \bigotimes_{j=0}^{k-1} \sH_{2j}$ and
$\sH_{\rm{out}_k} = \bigotimes_{j=0}^{k-1} \sH_{2j+1}$,
 is the comb of the reduced circuit $\mathcal C^{(k)}$ obtained by 
discarding the last $N-k$ teeth.

The connection of two circuit boards is represented by the {\em link
  product} of the corresponding combs $C_1$ and $C_2$, which is
defined as
  $C_1 * C_2 =\Tr_{\sK}[C_1^{\theta_{\sK}} C_2]$,
$\theta_{\sK}$ denoting partial transposition over the Hilbert space
$\sK$ of the connected systems (we identify with the same
labels the Hilbert spaces of connected systems).


One can wonder whether each positive operator which satisfies Eq. (\ref{recnorm})
corresponds to a sequential network of quantum channels. The answer is indeed positive, as shown in Refs. \cite{watrousgame,architecture,comblong} with the following Stinespring dilation theorem: 
\begin{theorem} \label{dilationtheorem}
Let $C^{(N)}$ be a positive operator on $\sH_{\rm{out}}\otimes \sH_{\rm{in}}$, with $\sH_{\rm{in}} := \bigotimes_{j =
  0}^{N-1} \sH_{2j}$ and $\sH_{\rm{out}} :=
\bigotimes_{j=0}^{N-1} \sH_{2j+1}$. If $C^{(N)}$ satisfies Eq. \ref{recnorm},
 then it is the Choi-Jamio\l kowski operator of a sequential network  given by the concatenation of $N$ isometries: for every state $\rho \in {\rm Lin} (\sH_{\rm in})$ one has
\begin{equation}\label{concatenation}
\mathcal{C}^{(N)} (\rho) =    \Tr_{A_N}[V^{(N)}\cdots V^{(1)} \rho V^{(1)\dagger} \cdots V^{(N)\dagger}]
\end{equation}
where $V^{(k)}$ is an isometry from $\hilb{H}_{2k-2}\otimes \hilb{H}_{A_{k-1}}$ to  $\hilb{H}_{2k-1} \otimes \hilb{H}_{A_{k}}$, 
and $\hilb{H}_{A_k}$ is an ancillary space, $\hilb{H}_{A_0}= \mathbb{C}$ (in Eq. (\ref{concatenation}) we omitted the identity operators on the Hilbert spaces where the isometries do not act).
\end{theorem}
This result, however, provides little insight on how to construct the isometries.    We now give the explicit construction in terms of the Choi-Jamio\l kowski operator in a way that can be automatically evaluated by a computer routine:
\begin{theorem}\label{decompolemma}
The minimal dimension of the ancilla space $\hilb {H}_{A_k}$ in Theorem \ref{dilationtheorem}  is the dimension of the support of $C^{(k)}$.  A choice of isometries $V^{(k)} :  \hilb{H}_{2k-2}\otimes \hilb{H}_{A_{k-1}} \to \hilb{H}_{2k-1} \otimes \hilb{H}_{A_{k}}$ with minimal ancilla space is obtained by taking $\hilb{H}_{A_k} = \supp(C^{(k)*})$, where $*$ denotes the complex conjugation in the canonical basis, 
and  
\begin{align}\label{eq:innerisometry}
  V^{(k)} = &
 I_{2k-1} \otimes  C^{(k)\frac12*} C^{(k-1)-\frac12 *} \,\,\times \nonumber \\
& \ \Ket{I}_{(2k-1)(2k-1)'}  T_{(2k-2)\rightarrow (2k-2)'}
\end{align}
where $T_{n \rightarrow m}=\sum_i\ket{i}_m\bra{i}_n $.
\end{theorem}
\Proof One  has $V^{(k) \dag}  V^{(k)} =     \left( C^{(k-1)*}\right)^{-\frac 12} \Tr_{2k-1}  [  C^{(k)^*}] \left( C^{(k-1)*}\right)^{-\frac 12} $, and Eq. (\ref{recnorm})  yields  $V^{(k) \dag}  V^{(k)} = I_{2k-2} \otimes I_{\supp (C^{(k-1)*})} =   I_{2k-2} \otimes I_{A_{k-1}}$.  Therefore, $V^{(k)}$ is an isometry.   Now, define the isometry $W^{(k)}= V^{(k)} \dots V^{(1)}$, which goes from $\hilb H_{{\rm in}_k}$ to $\hilb H_{{\rm out}_k} \otimes \hilb H_{A_k}$. 
By definition one has $W^{(k)} =  \left[I_{{\rm out}_k} \otimes   \left(  C^{(k)*}\right)^{\frac 12} \right] 
[|I \kk_{({\rm out}_k) \, ({\rm out}_k)'} \otimes T_{{\rm in}_k \rightarrow ({\rm in}_k)'}]$. However, according to Ref. \cite{JMP}, this is the minimal isometry of the channel $\mathcal C^{(k)}$.  Since the isometry is minimal, it is not possible to choose an ancillary space smaller than $\hilb H_{A_k}$. Finally, since $\mathcal C^{(N)}$ is nothing but the channel associated to the network, Eq. (\ref{concatenation}) follows. \qed

Theorem \ref{decompolemma}  implies  Theorem \ref{dilationtheorem},
and provides a recipe for the concrete realization of the quantum network with minimal
dimension of the ancillas at each step.
The dimension of the ancilla is the quantum``space'' of the computational network. Note that sometimes the isometries $V^{(k)}$ can act trivially on some subsystem,
this resulting in further simplifications of the physical implementation.  



\begin{figure}[t]
  \includegraphics[width=\columnwidth]{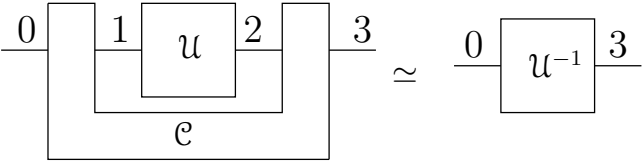}
  \caption{\label{figinversion}.
The quantum circuit $\mathscr{C}$, when linked with  the unitary channel 
$\mathscr{U}:{\rm Lin}(\mathcal{H}_1) \rightarrow {\rm Lin}(\mathcal{H}_2) $,  tries to reproduce
the action of  $\mathscr{U}^{-1}$ from input $\mathcal{H}_0$ to output $\mathcal{H}_3$.
}
\end{figure}

As an application of the methods outlined above we now consider the problem of finding
 the quantum network
that realizes the optimal inversion of a unitary operation.
Such a network consists of a circuit board $\mathscr C$ with an empty
slot to be linked to the unitary channel 
$\mathscr U (\rho)= U \rho U^\dag$ sending states on $\sH_1$ to states on $\sH_2$.
 The resulting circuit
$\mathscr C * \mathscr U$ has to be as  similar as possible to the channel
$\mathscr U^{-1}$ (see Fig. (\ref{figinversion})).
The quantum comb of $\mathscr C$ is $C \in {\rm Lin}(\hilb{H}_{3210})$, with $\sH_3 \simeq \sH_2 \simeq \sH_1 \simeq \sH_0\simeq \mathbb C^d$, and, according to Eq. (\ref{recnorm}), satisfies the normalization
\begin{equation}\label{norminverter}
\Tr_3[C] = I_2 \otimes C^{(1)}, \qquad   \Tr_1[C^{(1)}] = I_0.
\end{equation}
 Choi operator  of the unitary channel is $\KetBra{U}{U}_{21}$
and the link $\mathscr C * \mathscr U$ gives the operator $\Bra{U^*}_{21}C\Ket{U^*}_{21}\in {\rm Lin} (\hilb H_{30})$.
To quantify the similarity between the channel
$\mathscr C * \mathscr U$ and the target $\mathscr U^{-1}$ we use the channel fidelity
\cite{raginsky}:
if $\mathcal{A}$ and ${\mathcal B}$ are two channels and $A$ and $B$ are their  Choi-Jamio\l kowski operators
the channel fidelity $\mathcal{F}(\mathcal{A},{\mathcal B})$ is defined as $f(d^{-1}A,d^{-1} B)$ where $f$ is
the state fidelity $f(\rho, \sigma) = |\Tr\sqrt{\sqrt{\sigma}\rho\sqrt{\sigma}}|^2$.
In our case we have
\begin{align}
F(\mathcal C,\mathcal U) & = f(d^{-1}(C *\KetBra{U}{U}_{21}),d^{-1}\KetBra{U^\dagger}{U^\dagger}_{30})  &\nonumber \\
& = \frac{1}{d^2}  \Bra{U^\dagger}_{30}\Bra{U^*}_{21} C \Ket{U^\dagger}_{30}\Ket{U^*}_{21}.&
\end{align}
We assume the unknown unitary $U$ randomly  distributed according to the Haar measure of $SU(d)$, and, as a figure of merit, we adopt the average of the gate fidelity:
\begin{align}\label{averfidelity}
\overline F &= \int_{SU(d)} \!\!\!  dU F(\mathcal C,\mathcal U) & \nonumber \\
 &=\frac{1}{d^2}  \int_{SU(d)} \!\!\! dU \Bra{U^\dagger}_{30}\Bra{U^*}_{21} C \Ket{U^\dagger}_{30}\Ket{U^*}_{21}&
\end{align}
where $dU$ denotes the invariant Haar measure.
The following lemma holds:
\begin{lemma}
The operator $C$  maximizing the fidelity
  (\ref{averfidelity}) can be assumed without loss of generality to satisfy
  the commutation relation
\begin{equation}
[C, U_3 \otimes W_2 \otimes U_1\otimes W_0]=0 \quad \forall V,W \in SU (d)~. \label{CovR}
\end{equation}  
\end{lemma}
The proof consists in the standard averaging argument:
Let $C$ be optimal. Then take its average
$\overline{C} = \int \d U \d W~  (U_3 \otimes
 W_2 \otimes  U_1 \otimes  W_0) C (U_3 \otimes
 W_2 \otimes  U_1 \otimes  W_0)^{\dagger}$: it is immediate to see that
$\overline {C}$ satisfies Eqs. (\ref{CovR}) and
(\ref{norminverter}), and has the same fidelity as $C$.

Thanks to Schur's lemmas $C$ can be decomposed as
\begin{align}\label{decomb}
C = \sum_{\mu, \nu \in \mathsf{S}}a^{\mu \nu}P^{\mu}_{31} \otimes P^{\nu}_{20},
\end{align}
where $\mathsf{S} = \{ +,- \}$, $P^{\pm}_{ij}$ is the projector onto the symmetric/antisymmetric subspace of
$\mathcal{H}_i\otimes \mathcal{H}_j$ ,  and $a^{\mu \nu} \geq 0$ $\forall \mu, \nu$.
Moreover, using Eq. (\ref{decomb}) the fidelity (\ref{averfidelity}) becomes
\begin{align}
\overline F  &= \frac1{d^2} \Bra{I}_{30}\Bra{I}_{21} C \Ket{I}_{30}\Ket{I}_{21} & \nonumber \\
   &  =\frac{1}{d^2}\sum_{\nu \in \mathsf{S}} a^{\nu\nu }d_\nu , \; \quad d_\nu = \Tr[P^\nu],
\end{align}
while the normalization (\ref{norminverter}) becomes
$\sum_{\mu \in \mathsf{S}}a^{\mu\nu}d_\mu = 1, \forall \nu\in \mathsf S$.
The last equality implies the  bound
$\overline F = \frac{1}{d^2}\sum_{\mu \in \mathsf{S}} a^{\mu\mu }d_\mu \leq  2/d^2$,
which is achieved if and only if $a^{\mu\nu }= \frac{\delta_{\mu \nu}}{d_\mu}$,  that is, if and only if
\begin{align}\label{optimalinverter}
C&= \frac{P_{31}^+\otimes P_{20}^+}{d_+}+\frac {P_{31}^-\otimes P_{20}^-}{d_-} \nonumber\\
&  =\int_{SU(d)}  d \hat U  ~ |\hat U^\dag\kk\bb \hat U^\dag|_{30}  \otimes |\hat U^*\kk \bb \hat U^*|_{21}.
\end{align}
We now use Theorem \ref{decompolemma}
 to construct the optimal network from the quantum comb $C$.
Since $C^{(1)}=d^{-1}I_{10}$ the first isometry is given by
\begin{align}\nonumber
  V^{(1)} = \left( I_1 \otimes { C^{(1) *}}^{\frac12} \right) \Ket{I}_{11'}\otimes T_{0\to0'}=
 \frac{1}{\sqrt{d}} \Ket{I}_{11'}\otimes T_{0\to 0'},
\end{align}
namely it consists in the preparation of the maximally entangled state $\frac{1}{\sqrt{d}}\Ket{I}_{11'}$ while the input state is stored in a subsystem of  the ancilla space $\hilb H_{A_1}  \subset \hilb H_{1'0'}$.

The second isometry $V^{(2)}:\hilb{H}_2 \otimes \hilb H_{A_1}  \rightarrow \hilb{H}_{3} \otimes \hilb H_{A_2}$ is given by
\begin{align}\label{secondisometry}
V^{(2)}= (\sqrt{d}I_{3}\otimes C^{*\frac12}) \Ket{I}_{33'} \otimes T_{2\to2'}.
\end{align}
Remarkably, this is the Stinespring isometry of a measure-and-prepare channel. Indeed, consider  the channel $\mathcal{E} (\rho) =\Tr_{A_2}[V^{(2)}\rho V^{(2)\dagger}]$ and the POVM   
\begin{equation} Q_{\hat U} =  (C^*)^{-\frac 12}   |\hat U^T\kk\bb \hat U^T|_{3'0'}  \otimes |\hat U\kk \bb \hat U|_{2'1'} (C^*)^{-\frac 12},  
\end{equation}  
which provides a resolution of the identity in  $\sH_{A_2} = \supp  (C^*)$ due to Eq. (\ref{optimalinverter}). We then have
\begin{align}
\mathcal{E}(\rho) &= \int d\hat U \Tr_{A_2}[V^{(2)}\rho V^{(2)\dagger} Q_{\hat U}] 
 \nonumber \\
&= d\int d\hat U     U^\dag  \bb U|_{2'1'} \rho \Ket{U}_{2'1'} U,
\end{align}
namely the channel $\mathcal E$ can be implemented by measuring the POVM $P_{\hat U} = d |U\kk\bb U|_{2'1'}$ on the Hilbert space $\sH_{2'1'}$ and subsequently performing the unitary $\hat U^{\dag}$ on $\sH_{0'}$.   Therefore, the optimal network for the inversion of an unknown unitary channel corresponds to an``estimate and re-prepare" strategy:
first the isometry $V^{(1)}$ provides the optimal input for the estimation of $U$ (that is, the maximally entangled state $d^{-\frac 12}|I\kk_{11'}$), then, after the unknown unitary has been applied, 
the second channel $\mathcal E$ performs the optimal POVM on the state $d^{- \frac 12} \Ket{U}_{11'}$ and, depending on the estimate
$\hat U$, applies the unitary $\hat U^\dagger$ on the input state stored in wire $0'$. The physical implementation involving measurement and classical feed-forward is an alternative to the coherent, fully quantum processing corresponding to the isometry $V^{(2)}$.

\begin{figure}[t]
  \includegraphics[width=\columnwidth]{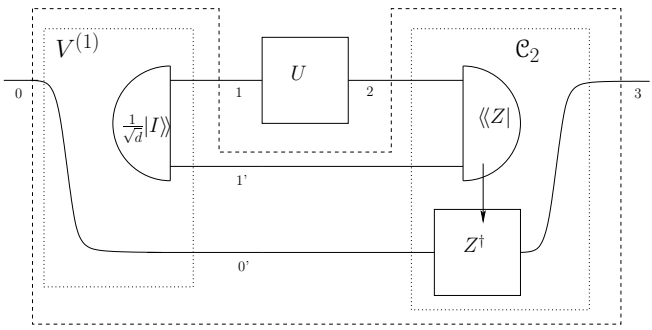}
  \caption{\label{figoptinverter}.
Optimal circuit for the inversion of a unitary transformation.
the input state in wire $0$ is stored in a quantum memory. The unitary $U$ to be inverted
is estimated and the inverted estimated unitary is applied to the input.}
\end{figure}

In conclusion, we provided a general method for recovering all the
isometries of a network from its Choi-Jamio\l
kowsky operator minimizing the comptational space. This result allows us to formulate an algorithm
for designing optimal quantum networks for any desired task (e. g. cloning,
discrimination, estimation): 
\begin{enumerate}
\item Choose a suitable figure of merit $F$ for the task of interest.
\item  Find the positive operator $C$ satisfying constraint
  in Eq.~(\ref{recnorm}) and maximizing $F$.
\item Set $C^{(0)}=1$ and $I_{A_0} = 1$.
\item For $k=1$ to $k=N$ do the following
  \begin{enumerate}
  \item Calculate $I_{\overline{\rm{in}_k}}\otimes C^{(k)} = \Tr_{ \overline{\rm{out}_{k} }}[C]$, where $I_{\overline \sH}$  ($\Tr_{\overline \sH})$ denotes the identity (partial trace) over all Hilbert spaces but $\sH$  
  \item Define $V^{(k)}$ as in Theorem \ref{decompolemma}
  \end{enumerate}
\item The optimal network is given by the concatenation of the
    $V^{(k)}$'s in Eq. (\ref{concatenation})
\end{enumerate}
We applied the algorithm to design the optimal circuit for the
inversion of a unitary transformation.
It is worth noting that in general the numerical optimization of step 2 can be challenging, and that it is typically convenient to exploit the symmetries of the problem to reduce the number of parameters, as we did here in our example. On the other hand, the remaining steps 3-5---which
represent the original result of the present Letter---can be easily programmed on a computer.

\par {\em Acknowledgments.---} 
This work is supported by
Italian Ministry
of Education through grant PRIN 2008 and
 the EC through  project COQUIT.
Research at
Perimeter Institute for Theoretical Physics is supported
in part by the Government of Canada through NSERC
and by the Province of Ontario through MRI.



\end{document}